\documentclass[12pt,preprint]{aastex}

\newcommand\um{\ensuremath{\mathrm{\ \mu{m}}}}

\newcommand\cm{\ensuremath{\mathrm{\ cm}}}
\newcommand\pc{\ensuremath{\mathrm{\ pc}}}
\newcommand\yr{\ensuremath{\mathrm{\ year}}}
\newcommand{\LkHa}{LkH$\alpha$ }

\shorttitle{Edge-on Disk around PDS 144N}
\shortauthors{Perrin et al.}

\slugcomment{Accepted for publication in ApJ}

\begin{document}

\title{Discovery of an Optically Thick, Edge-on Disk \\
around the Herbig Ae star PDS 144N\footnote{Some of the data
presented herein were obtained at the W.M. Keck Observatory, which is
operated as a scientific partnership among the California Institute of
Technology, the University of California and the National Aeronautics
and Space Administration. The Observatory was made possible by the
generous financial support of the W.M. Keck Foundation.}}

\author{Marshall D. Perrin}
\affil{Astronomy Department, University of California,
    Berkeley, CA 94720-3411}
\email{mperrin@astro.berkeley.edu}

\author{Gaspard Duch\^ene}
\affil{ Laboratoire d'Astrophysique, Observatoire de Grenoble, BP 53, F-38041
Grenoble Cedex 9, France}

\author{Paul Kalas, James R. Graham}
\affil{Astronomy Department, University of California,
    Berkeley, CA 94720-3411}

\begin{abstract}

We have discovered an optically thick, edge-on circumstellar disk
around a Herbig Ae star in the binary system PDS 144, providing the
first intermediate-mass analog of HK Tau and similar T Tauris.  This
system consists of a $V\sim 13$ mag. primary and a fainter companion,
with spectra of both stars showing evidence for circumstellar disks
and accretion; both stars were classified as Herbig Aes by the Pico
dos Dias survey.  In Lick adaptive optics polarimetry, we resolved
extended polarized light scattered from dust around the northern star.
Followup Keck adaptive optics and mid-infrared observations show that
this star is entirely hidden by an optically thick disk at all
wavelengths from 1.2 to 11.7 microns. The disk major axis subtends
$\sim0\farcs8$ on the sky, corresponding to $\sim 800$ AU at a
distance of 1000 pc. Bright ``wings'' extend $0\farcs3$ above and
below the disk ansae, due most likely to scattering from the edges of
an outflow cavity in a circumstellar envelope. We discuss the
morphology of the disk and the spectral energy distributions of the
two PDS 144 stars, present preliminary disk models, and identify a
number of open questions regarding this fascinating system.

\end{abstract}


\section{Introduction}

Circumstellar disks inescapably form around young stars, as angular
momentum conservation prevents infalling matter from accreting directly onto the
stellar surface.  Transfer of angular momentum between different parcels
of matter in these disks not only allows continued accretion, but has been
implicated in the launching of spectacular bipolar jets and winds
\citep{1994ApJ...429..781S}. 
Studying the
detailed physics of such disks is thus important for understanding the
nature of these outflows, which provide feedback into the star
formation process and help limit the star formation rate.   
The standard model for T
Tauri stars invokes magnetic fields to drive the outflows. While strong magnetic fields are
inevitable in convective, low mass stars, the intermediate mass Herbig
Ae/Be stars \citep{her60,1998ARA&A..36..233W} 
have much shallower convective zones and consequently have much
weaker dynamos. Thus it is natural to suppose that disk environments
around these stars may differ substantially from those around T Tauri
stars.

Characterizing these disks also illuminates the conditions in
which planet formation takes place.
The tremendous diversity of
recently discovered extrasolar planetary systems raises the question
of whether this diversity arises from widely differing initial
conditions in circumstellar disks, or if instead all solar systems begin
similarly but evolve apart via planetary dynamics.  Are
conditions around A- and B-type stars as conducive to
planet formation as are those around lower-mass stars? 
 Present Doppler
 planet searches generally exclude A and B stars due to their relatively
 featureless spectra (though see \citet{2005A&A...443..337G} for an exception to this rule), yet the presence of debris disks around the A
 stars Vega \citep{1984ApJ...278L..23A} and $\beta$ Pic
 \citep{1984Sci...226.1421S} provided the first evidence for the
 existence of extrasolar planetary systems. Indeed, more recent observations
 of perturbations to the disks of Vega and Fomalhaut
 indicate the
 presence of Jupiter-mass planets around those stars
 \citep{2002ApJ...569L.115W,Kalas2005}.
 The Herbig Ae/Be stars \citep{her60}
are the progenitors of Vega-like debris disks. By studying the
circumstellar
material around these stars, we aim to quantify the evolution of dust
properties and spatial distribution with stellar age, and thereby gain
insight into the growth of larger dust particles and ultimately
planets.

Optically thick, edge-on disks are particularly amenable to spatially
resolved observations in the visible or infrared.  Disk imaging is
frequently complicated by point spread function (PSF) subtraction artifacts, especially on
small angular scales.  For optically thick, edge-on disks, the lack of
direct starlight greatly reduces the dynamic range needed to image
the disk.  At the same time, modeling the observed light distribution
can constrain the disk's physical parameters, such
as the degree of flaring and dust grain growth
\citep{2001ApJ...553..321D,2004ApJ...602..860W}.
Edge-on disks are particularly valuable for such studies, as
the disk's vertical structure is directly apparent. By modeling  
multiwavelength observations
of edge-on disks, one can determine the wavelength dependence of 
the dust scattering opacity, which provides 
insight into dust grain properties \citep{cotera2001,2004ApJ...602..860W}.
Edge-on disks in
binary systems may additionally shed light on
aspects of the binary formation process and disk/disk interactions.

Edge-on disks are known around a number of lower-mass T Tauri stars,
most notably
\object{HH 30} \citep{1996ApJ...473..437B}, \object{HK Tau} B
\citep{1998ApJ...502L..65S}, and \object{HV Tau} C \citep{2000A&A...356L..75M}, but until 
now no edge-on disk has been seen around a Herbig Ae or Be star.
We present in this paper our discovery of such an edge-on circumstellar
disk around the Herbig Ae star \object{PDS 144N} ($\alpha =
15^h49^m15.5^s$, $\delta = -26\degr00'50.2"$, J2000). 

\citet{carballo1992} identified IRAS 15462-2551 with a previously
anonymous 12th magnitude star in Scorpius. The Pico dos Dias Survey, a
spectroscopic followup to IRAS sources \citep{torres1995},
subsequently named this source PDS 144, discovered it to be a
$5\arcsec$ binary, and classified both members as Herbig Ae stars.
\citet{vieira2003} report a spectral type of A2IV for the northern
member and A5V for the southern.  We follow their notation in
referring to these stars as ``PDS 144N'' and ``PDS 144S'',
respectively.  In addition to the H$\alpha$ emission signature of
accretion, they found [SII] $\lambda\lambda$ 6717, 6730  and [OII]
$\lambda\lambda$ 3726, 3728 emission around both stars, indicative of
a photodissociation region. These
lines are good tracers of outflows around other young stars, and their
presence makes it probable that such outflows are present around both
PDS 144 stars. 
The physical association of these two stars has not yet been
rigorously shown, but given their proximity and the fact that both are 
Herbig Ae stars which are currently accreting and thus of similar age, 
it is very likely that they are indeed bound and coeval.

We describe our observations in \S 2, and discuss the 
distance to PDS 144 in \S 3.
In \S 4, we examine the observed disk's morphology, 
before presenting preliminary models for the disk in \S 5.
Lastly, \S 6 recapitulates our conclusions and identifies a
number of remaining open questions about this intriguing system.

\section{Observations}

PDS 144N's disk was discovered in observations
obtained with Lick Observatory's 3-m Shane telescope, and
was subsequently imaged with both 10-m W. M. Keck telescopes. 
Together, these observations span the wavelength range 1.2 to 11.7
microns.  An observation log is presented in Table \ref{obslog}, while Table \ref{fluxtable} gives our measured photometry for both PDS 144 stars.

\begin{deluxetable}{llllrl}
\tablecolumns{4}
\tablewidth{0pt}
\tablecaption{\label{obslog}Observation Log}
\tablehead{
\colhead {Telescope} & \colhead{Instrument} & \colhead{Filter} & \colhead{Date} & \colhead{$t_{int}$ (s) } & {$\theta$ (\arcsec)}
}
\startdata
Lick 3 m & IRCAL & $J$ & 2004 July 5 &  320  &  0.38  \\
\nodata &\nodata & $H$ & 2004 July 5 & 480   &  0.33  \\
\nodata &\nodata & $K_s$ & 2004 July 5 & 240 &  0.30  \\
Keck II & NIRC2 & $H$ & 2004 Aug 29,30 & 120 &  0.055  \\
\nodata &\nodata & $K'$ & 2004 Aug 29 & 90   &  0.058  \\
\nodata &\nodata & PAH & 2004 Aug 30 &  144  &  0.07  \\
\nodata &\nodata & $L'$ & 2004 Aug 30 & 90   &  0.08  \\
Keck I  & LWS    & $11.7\um$ & 2004 Aug 28 & 85 & 0.35  \\
\enddata
\tablecomments{Observing Log for PDS 144. $t_{int}$ is the total integration time in each filter, and $\theta$ is the
achieved resolution in arcseconds, measured from a point source.
See Table 2 for a detailed description of filter properties.}
\end{deluxetable}

\subsection{Lick}
We first observed PDS 144 as part of our ongoing adaptive optics polarimetry
survey of Herbig Ae/Be stars \citep{Perrin2004Sci}, being carried out with the
Lick Observatory Adaptive Optics (AO) system \citep{2002SPIE.4494..336G} and the IRCAL
camera \citep{Lloyd00}.  IRCAL's differential polarimetry mode incorporates a
Wollaston prism and rotating half-wave plate to obtain high-contrast images in
polarized light \citep{Perrin2004SPIE}.  For these observations, the AO system
was locked on PDS 144S. The seeing was excellent (Fried's
parameter $r_0 \sim 16 \cm$ at 550 nm), but the target was observed at an airmass of
2.25, and AO correction was limited to a Strehl ratio of 0.1 at $K_s$
band. After dark subtraction and flat fielding, the polarimetric data were
reduced via a double-differencing code based on the algorithm of \citet{kuhn01},
resulting in images of the Stokes parameters $I$, $Q$, and $U$ for each near
infrared band.  These data were then flux calibrated using the standards \object{HD
161903} \citep{EliasStandards} and [PMK98] 9170 \citep{PerssonStandards}.

\begin{figure}[!ht]
\begin{center}
\plotone{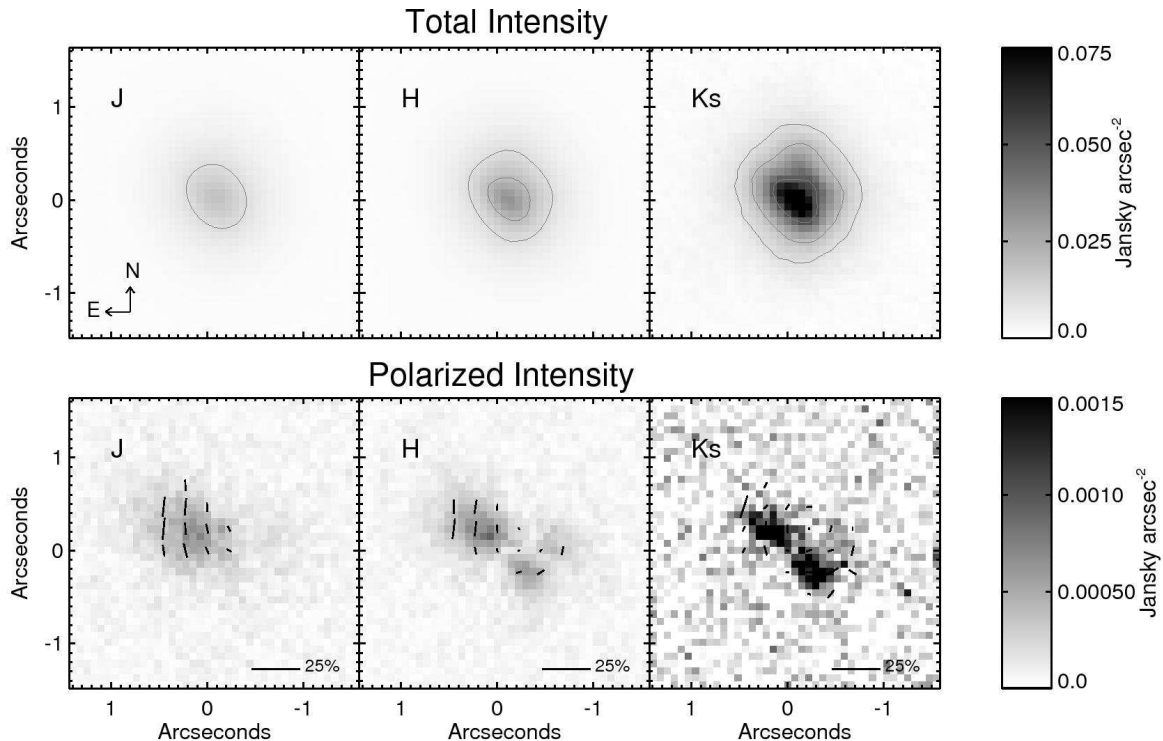}
\end{center}
\vskip -0.30in
\caption{
\footnotesize
\label{lickfig}
Lick AO polarimetry observations of PDS 144N. The top row
shows total intensity, $I$, while the second row shows
polarized intensity, $\sqrt{Q^2+U^2}$, with position vectors showing the
orientation and magnitude of polarization. The color scales are the same for
each row, as shown at right. Contours are at 0.01, 0.02, and 0.05 Jy arcsec$^{-2}$. 
In total intensity, the morphology is clearest in $K_s$ band, where the higher Strehl 
ratio PSF resolves a dark lane crossing PDS 144N. In polarized intensity,
a bright bar of nebulosity is visible at a position corresponding to the
southeast side of the dark lane--presumably the disk face inclined toward us. A
portion of the northwest face is visible in the $K_s$ and $H$ bands, while in
$J$ only the far eastern end of the nebula is detected. The differences between
wavelengths may be due to obscuring foreground material, or may be artifacts of the
relatively low Strehl ratio (3\%) of our $J$ band observations. 
These patterns of position vectors are consistent with
light scattered from circumstellar material around PDS 144N.
}
\end{figure}

The northern source is resolved at all wavelengths, while the southern is
point-like (Figure \ref{lickfig}). At $H$ and especially $K_s$ band, PDS 144N
appears as an elliptical nebula bisected by a dark lane.  In polarized light,
PDS 144N is extended $\sim 1\arcsec$ from northeast to southwest, with polarization
position vectors roughly perpendicular to this.  Centrosymmetric polarization
vectors are the signature of scattered starlight, and the
presence of such vectors along one axis is precisely what is expected for
an edge-on circumstellar disk.  The observed polarization fractions range between 5-10\%, though this is a lower
limit for the true polarization since the observed values are biased downward by
AO-uncorrected light.

In polarized light, PDS 144S appears only at the 1-2\% level of systematic
subtraction residuals expected for an unpolarized source. The position vectors
are not centrosymmetric, and based on these polarization data alone there is no
direct evidence for a disk around PDS 144S.

\clearpage
\subsection{Keck}

\begin{figure}[!ht]
\begin{center}
\plotone{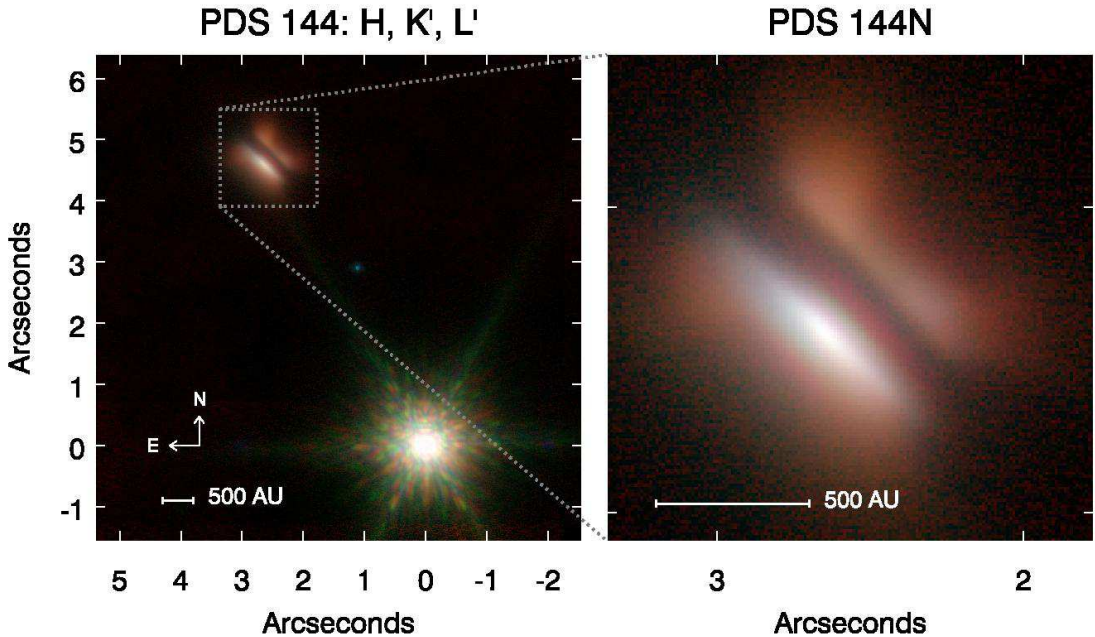}
\end{center}
\vskip -0.30in
\caption{
\footnotesize
\label{keckfig}
Keck AO imaging of PDS 144. The $H, K'$ and $L'$ bands are shown
as blue, green, and red respectively. PDS 144N appears as an edge-on disk seen
in silhouette against brighter nebulosity, which extends vertically on both
sides of the disk plane. Both PDS 144N and the brighter PDS 144S are Herbig Ae
stars, while the nature of the faint third source located almost directly
between them is unknown. 
The scale bar in AU assumes a distance of 1000 pc; see section \ref{distance}.
The nebulosity around PDS 144N flares vertically near its edges, reminiscent of
a pair of opposing wingnuts. 
In the northwestern, fainter face of the nebula, the wings are
sufficiently prominent that the nebula is brighter toward its outer
edges than in the middle: there is a local minimum in intensity on the
symmetry axis. 
The presence of a circumstellar envelope in
addition to a disk is required to explain these ``wings''.  
}

\end{figure}

\begin{figure}[!ht]
\begin{center}
\plotone{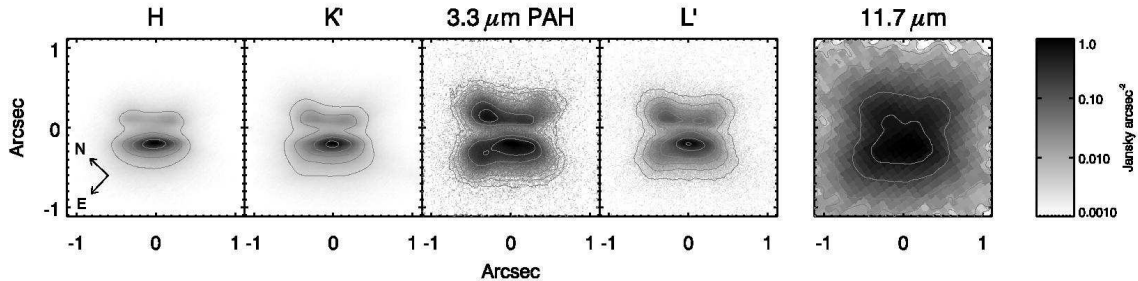}
\end{center}
\vskip -0.30in
\caption{
\footnotesize
\label{diskfig}
The PDS 144N disk, as observed with NIRC2 on Keck II (left four
frames) and LWS on Keck I (right frame). All are displayed with the same
logarithmic intensity scale, as indicated at right. Contours are at
0.01, 0.025, 0.05, 0.1, 0.25, and 0.5 Jy arcsec$^{-2}$.
The vertical extent of the dark lane and the overall width of the nebulosity do
not change significantly across the four NIR filters, but the ``wings'' of
the nebula become more prominent in $L'$ and especially the 3.3 $\micron$ PAH
filter.  The relative brightness of the 3.3 $\micron$ image indicates that PAH
grains must be present around PDS 144N. 
In the 11.7 $\micron$ filter, the dark lane is no longer visible as a local
minimum, but appears as a broad plateau to one side of the bright southeast
disk face. 
The presence of PAHs complicates the interpretation of the 11.7 $\micron$ image: 
The light we detect at this wavelength may be scattered thermal emission from
the hot inner disk, fluorescent emission from the 11.3 $\micron$ PAH
feature, which lies within the filter bandpass, or most likely a mix of the two.
}
\end{figure}

Shortly after the disk was discovered at Lick, we obtained additional
near-infrared data with the Keck II AO system
\citep{Wizinowich2000} and NIRC2 camera (Matthews, K., in
preparation).  PDS 144S again served as the wavefront reference. The
Keck images have much superior
AO correction than the Lick data, in part because PDS 144 is
observable at a more favorable airmass from Keck (Figs.
\ref{keckfig} and \ref{diskfig}).
Because the exposure times were chosen to maximize SNR on the PDS 144N
disk, PDS 144S is saturated in all wavelengths other than $K'$.  These
imaging data were reduced via the usual steps of dark and sky
subtraction, flat fielding, and mosaicing, and were flux-calibrated
using the same photometric standard, [PMK98] 9170. For the 3.3 $\micron$ PAH filter, no photometric
standard observations were available nor
was a typical zeropoint available from the NIRC2 documentation. We
therefore estimated a zeropoint for the PAH filter relative to our
measured zeropoints for the other filters, using the known PAH filter
response, detector quantum efficiency, and the typical Mauna Kea
atmospheric transmission across the filter bandpass \citep{Roethesis}.
To evaluate the reliability of this synthetic PAH zeropoint, we
similarly estimated a $K'$ zeropoint; the estimated and measured $K'$
zeropoints agreed within 0.05 mag.

The center of PDS 144N is located $5\farcs40\pm0\farcs01$ from PDS 144S,
at a position angle of $28.7\pm0.3\degr$. It is difficult to define an
exact center for an extended, obscured object; 
neither the location of peak brightness nor the photocenter are
appropriate, so we used a position roughly in the middle of the dark
lane but displaced slightly toward the brighter face. Based on our
disk models discussed below, this is a reasonable estimate for the
location of the hidden star.

In addition to the two known members of PDS 144, we detect three other
point sources within our full field of view mosaics (roughly
13\arcsec$\times$11\arcsec).  Photometry and astrometry for these
sources are given in Table \ref{othersources}.  Most notably, there is
a faint source located almost directly between the two stars of PDS
144. These are most likely field stars, either foreground or
background.  If so, and assuming they are not high proper motion
foreground objects, they may be useful as astrometric references for
PDS 144.  Its proper motion of 8 mas/year \citep{teixeira2000}
corresponds to one $K_s$-band Keck resolution element per four years,
so proper motion tests will be possible rapidly.

During the same observing run we also obtained a mid-infrared image
with the Long Wavelength Spectrometer (LWS;
\citet{jones_and_puetter1993}) on Keck I (Fig.  \ref{diskfig}).  LWS
was used with the 11.7 $\mu$m filter in chop-nod imaging mode with a
$10\arcsec$ chop throw at a position angle of $0\degr$ and a chop
frequency of 2.5 Hz. Data were reduced using the standard
double-difference subtraction to remove sky, telescope, and
instrumental backgrounds; no flat fielding was performed.  The
photometric standards \object{HD 197989} and \object{HD 2486}
\citep{CohenStandards} were used to flux-calibrate the data. 
We rotated and rebinned the LWS data to register it with our NIRC2 data, using
the documented plate scales of 9.94 and 84.7 mas/pixel respectively.
Image quality was near diffraction limited
(FWHM$\sim0\farcs35$). The LWS image shows very similar morphology to
the near-IR data, particularly if the near-IR data are reduced to
the same resolution as the LWS image. 

\subsection{Spitzer}

Spitzer observations of PDS 144 are available from the Spitzer Archive. These
observations, originally carried out as part of the "Molecular Cores
to Planets" legacy program \citep{evans_feps}, span 3.6-70 $\micron$
using IRAC and MIPS. We retrieved the pipeline-processed ``Post-BCD''
mosaics from the archive and performed photometry
with the APEX package \citep{2005PASP..117.1113M}. 
APEX was able to successfully
de-blend the sources, even with the substantial
PSF overlap at longer wavelengths. Since PDS 144 was the only source in the field, no
contemporaneous PSF measurement was available and so we used the reference
PSFs from the Spitzer web site. The Spitzer photometry is presented in Table \ref{fluxtable} and
figure \ref{sedfig}. 

We note in passing that the summed fluxes for PDS 144 in the MIPS 24 $\micron$
band are in excellent agreement with the 20 $\micron$ measurement by
IRAS, which did not resolve the two sources. However, the 60 $\micron$ IRAS flux is three times the MIPS 70
$\micron$ measurement, indicating that the IRAS measurement at that
wavelength suffers from confusion due to the large beam size. No
other bright sources are present in the MIPS field of view, so it is
unknown whether other source(s) or extended emission from diffuse
gas are responsible for this confusion.

\subsection{Spectral Energy Distribution}

\begin{figure}[!ht]
\begin{center}
\includegraphics[width=5.5in]{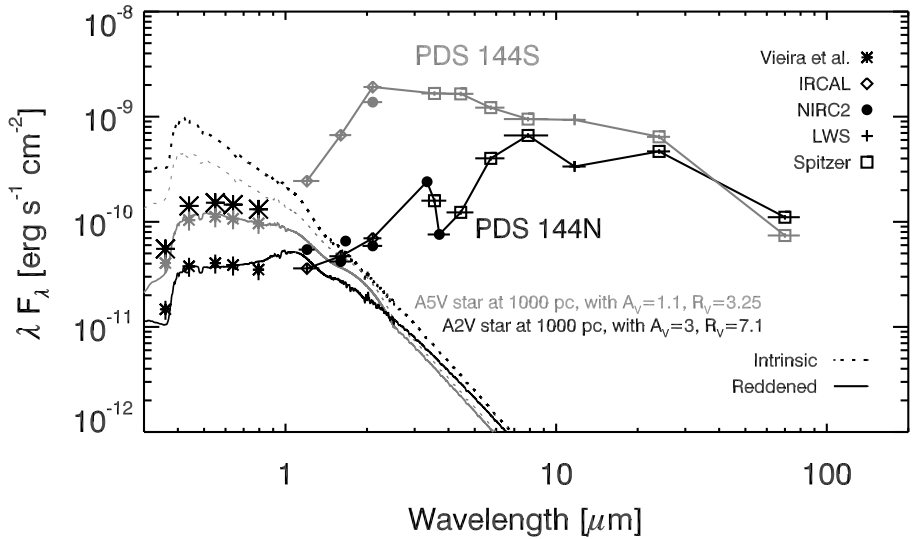}
\end{center}
\vskip -0.30in
\caption{
\footnotesize
\label{sedfig}
Spectral Energy Distribution for PDS 144. The key at right indicates
which instrument each plotted point is from.
Horizontal bars indicate the FWHMs of each filter bandpass, while 
vertical bars (typically smaller than the plot symbols) show the
uncertainties. 
For the 
optical points from \citet{vieira2003}, the larger asterisks (top) are their reported total
fluxes for both stars of PDS 144, while the two rows of smaller
symbols are our estimates for the two stars' individual fluxes, as
discussed in \S \ref{distance}.
We overplot both intrinsic and reddened stellar spectra for the two
stars. 
The A5 spectrum for PDS 144S has been attenuated by
$A_V=1.1, R_V=3.1$ using the extinction curve of
\citet{1989ApJ...345..245C}. Since PDS 144N is seen entirely in
scattered light at optical wavelengths, it is a crude approximation to
fit it with a reddened stellar spectrum. We show here
an A2 spectrum reddened by $A_V=3, R_V=7.1$ to demonstrate that the
observed optical magnitudes can be obtained with a moderate amount of
dust, and caution that $A_V=3$ does \textit{not} represent the direct
line-of-sight extinction to the star through the highly optically
thick disk midplane.
}

\end{figure}

The spectral energy distribution (SED) for the PDS 144 system is shown
in figure \ref{sedfig}.  In addition to our own measurements, we
plot the optical points from \citet{vieira2003}.  We 
overplot flux-calibrated stellar spectra corresponding to the spectral
types of PDS 144: A5 and A2 for PDS 144N and PDS 144S, respectively. 
We assume here a distance of 1 kpc, but see
\S 3 for a detailed discussion of the distance to PDS 144, which
remains very uncertain.

PDS 144S's SED rises sharply from 1 to 2 microns before leveling off,
while PDS 144N's SED is nearly flat throughout the optical and near
IR. They are both therefore Class II Herbig Ae stars in the
classification system of \citet{hil92}. Other Herbig Ae stars with
SEDs similar to PDS 144N include \LkHa 233 and Parsamian 21.  Those
sources are both embedded deeply enough that the stars are completely
hidden at optical and near-IR wavelengths and only scattered light is
seen \citep{1992ApJ...400..556S,Perrin2004Sci}, just as in PDS
144N.

At $70 \micron$, PDS 144N is brighter than its
southern neighbor. This suggests that while both stars possess
substantial circumstellar dust, PDS 144N has a greater amount of cool
dust at larger orbital radii than does PDS 144S.

\begin{deluxetable}{llrl|rr|rr}
\tabletypesize{\small}
\tablecolumns{4}
\tablewidth{0pt}
\tablecaption{\label{fluxtable}Measured Fluxes}
\tablehead{
Instrument	& Filter & $\lambda$ & FWHM      & PDS 144N & PDS 144S 	& PDS 144N & PDS 144S  \\
                &        & (\micron) & (\micron) & (Jy)     & (Jy)      & (mag.)   & (mag.) \\
}
\startdata
IRCAL	& $J$	& 1.238	& 0.267	& $0.014 \pm 0.002 $	& $  0.098 \pm 0.012 $	& $12.61 \pm 0.07 $	& $ 10.53 \pm 0.06 $\\
\nodata	& $H$	& 1.650	& 0.297	& $0.025 \pm 0.003 $	& $  0.357 \pm 0.026 $	& $11.52 \pm 0.06 $	& $  8.64 \pm 0.05 $\\
\nodata	& $K_s$	& 2.150	& 0.320	& $0.049 \pm 0.006 $	& $  1.343 \pm 0.102 $	& $10.31 \pm 0.10 $	& $  6.71 \pm 0.10 $\\
NIRC2	& $H$	& 1.633	& 0.296	& $0.023 \pm 0.003 $	& $> 0.228 \pm 0.031 $	& $11.64 \pm 0.02 $	& $< 9.15 \pm 0.02 $ \\
\nodata	& $K'$	& 2.124	& 0.351	& $0.043 \pm 0.004 $	& $  0.997 \pm 0.118 $	& $10.48 \pm 0.02 $	& $  7.06 \pm 0.02 $ \\
\nodata	& PAH	& 3.290	& 0.056	& $0.277 \pm 0.034 $	& $> 1.576 \pm 0.196 $	& $7.58 \pm 0.05 $	& $< 5.70 \pm 0.05 $ \\
\nodata	& $L'$	& 3.776	& 0.700	& $0.094 \pm 0.013 $	& $> 1.132 \pm 0.161 $	& $8.56 \pm 0.04 $	& $< 5.85 \pm 0.04 $ \\
LWS	& 11.7 & 11.698 & 1.104&  $1.352 \pm 0.249 $	& $  3.751 \pm 0.692 $	& $3.28 \pm 0.10 $	& $  2.17 \pm 0.10 $\\
IRAC    & 3.6  &  3.550 & 0.75 &  $0.188 \pm 0.010 $    & $  1.97~ \pm 0.099 $  & $7.94 \pm 0.06 $      &  $5.39 \pm 0.06 $     \\
\nodata & 4.5  &  4.439 & 1.01 &  $0.182 \pm 0.009 $    & $  2.44~ \pm 0.123 $  & $7.49 \pm 0.06 $      &  $4.67 \pm 0.06 $     \\
\nodata & 5.8  &  5.731 & 1.42 &  $0.767 \pm 0.039 $    & $  2.33~ \pm 0.118 $  & $5.44 \pm 0.06 $      &  $4.23 \pm 0.06 $     \\
\nodata & 8.0  &  7.872 & 2.93 &  $1.74~ \pm 0.089 $    & $  2.49~ \pm 0.127 $  & $3.92 \pm 0.06 $      &  $3.53 \pm 0.06 $     \\
MIPS    & 24   &  23.7  & 4.7  &  $3.74~ \pm 0.190 $    & $  5.15~ \pm 0.260 $  & $0.70 \pm 0.06 $      &  $0.35 \pm 0.06 $     \\
\nodata & 70   &  71.4  & 19   &  $2.58~ \pm 0.140 $    & $  1.73~ \pm 0.100 $  &$-1.31 \pm 0.06 $      & $-0.88 \pm 0.06 $     \\
\enddata
\tablecomments{Photometry for PDS 144. The IRCAL and NIRC2
measurements used a $1\arcsec$ radius aperture, while $1\farcs7$ was used for
LWS. Aperture correction factors were computed using the Keck AO PSF model
of Sheehy et al. (in preparation) and are typically 2-4\%. Flux calibrations are as discussed in the
text; the filter wavelengths listed above are isophotal wavelengths
computed according to the method of \citet{CohenStandards}. Columns marked with a $<$ or $>$ are limits derived from
saturated observations of PDS 144S. 
The flux uncertainties in Jy include contributions from both the aperture photometry
and the absolute flux calibration. For Spitzer data, filter properties are taken from \citet{2004ApJS..154...10F} and \citet{2004ApJS..154...25R}. 
Photometry and statistical uncertainties were computed with the APEX package, plus
an estimated 5\% absolute flux calibration error for all Spitzer
bands.
}
\end{deluxetable}

\begin{deluxetable}{lrrr}
\tablecolumns{3}
\tablewidth{0pt}
\tablecaption{Additional sources detected}
\tablehead{ 
\colhead{Source \#} & \colhead{R.A. Offset ($''$)} & \colhead{Dec. Offset ($''$)} & \colhead{$H$ Mag.} 
} 
\startdata
3 & 1.122 & 3.000 & $15.50 \pm 0.10$ \\
4 & 7.768 & 3.480 & $16.55 \pm 0.20$ \\
5 & 8.908 & 2.505 & $17.90 \pm 0.20$ \\
\enddata
\tablecomments{\label{othersources} R.A. and Dec. offsets are relative to PDS 144S, and have uncertainties of $\pm 0\farcs005$.}
\end{deluxetable}

\section{The Distance to PDS 144}
\label{distance}

The distance to PDS 144 remains frustratingly uncertain. Previous estimates 
in the literature 
range from 140 to 2000 parsecs. The nearer
distance was suggested by \citet{teixeira2000}, who
assumed that PDS 144 lies in the Upper Scorpius--Ophiucus star forming
region, based primarily on its position in the sky. 
The more distant estimate is from \citet{vieira2003}, who derived an optical
photometric distance of 2000 pc for PDS 144N, and 1030 pc for PDS 144S.

The lack of resolved optical photometry for PDS 144 complicates
the determination of a photometric distance.
Resolved optical photometry will be available from HST
observations scheduled for the near future. 
Until then, we must make do with a best estimate from the currently
available data. 
Our near-IR measurements are unsuitable 
for photometric distance measurements, since they are presumably contaminated by 
IR excess from circumstellar dust. We note that the 
$K'$ band flux of PDS 144S is consistent with the photospheric flux from an A5
star located at 140 pc.
Since the true $K'$ luminosity is presumably greater than photospheric, 
the actual distance must be $\geq 140$ pc.

 While \citet{vieira2003} only report combined
photometry for PDS 144 ($V=12.8$ mag), they measured $\Delta V = 1.1$
mag. from the relative counts of the two sources in their
spectroscopic observations (Carlos Torres, private communication).
This implies $V = 13.1$ and 14.2 for PDS 144S and 144N, respectively.
From these numbers, Vieira et al. derived their 1030 and 2000 pc
photometric distances.  Since we see PDS 144N only in scattered light,
it seems certain that its 2000 pc photometric distance is an
overestimate. But the photometric distance for the less-obscured PDS
144S may be accurate. Given PDS 144S's unresolved appearance,
it is plausible that we see its photosphere directly; certainly it is
not as obscured as PDS 144N. If we assume this is the case, then we
can follow \citet{vieira2003} and derive a photometric distance to PDS
144S.

\begin{figure}[!ht]
\begin{center}
\includegraphics[width=3in]{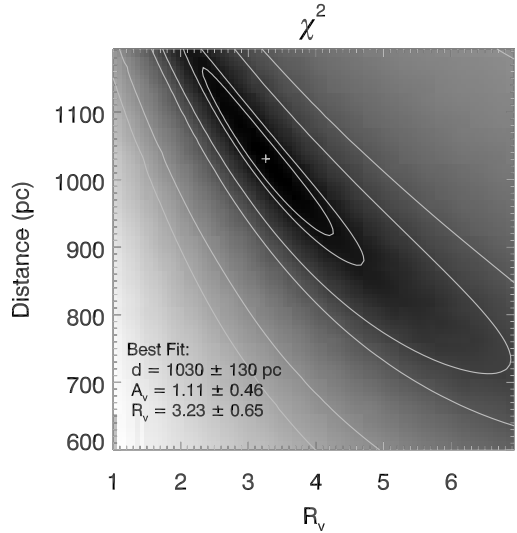}
\end{center}
\vskip -0.30in
\caption{
\footnotesize
\label{distfig}
We fit synthetic $UBVRI$ photometry to PDS 144S in order to estimate its distance, as described in the text, and plot here 
the resulting $\chi^2$ against distance and reddening.
Vieira et al. do not provide uncertainty
for their photometry; not knowing the error bars results in our
$\chi^2$ being uncertain by an arbitrary multiplicative constant. We
therefore normalize our best-fit $\chi^2=1$ and plot here contours of
0.5, 1, 5, 10, 50, and 100 above the best-fit $\chi^2$. Our best fit,
marked by a cross,
gives a distance of 1030 pc. While there is substantial
covariance between the distance and the extinction law parameters,
distances less than $\sim 500$ pc do not appear plausible.
}
\end{figure}

We estimated the optical fluxes of PDS 144S from the combined $UBVRI$ magnitudes 
and $\Delta V$ from Vieira et al., assuming that the flux ratio between the two sources
is the same for all wavelengths. (Though this assumption is crude, the following results are
only weakly dependent on it.)
We then performed a nonlinear least squares fit to PDS 144S using a flux-calibrated ZAMS $A5V$ 
stellar spectrum, reddened by dust
using the extinction law of \citet{1989ApJ...345..245C}
and then multiplied by appropriate filter bandpasses.  The free
parameters of the fit were the distance $d$, the visual extinction
$A_V$ and the reddening parameter $R_V$; we assume there is only one
dust component along the line of sight. The best fit parameters 
are $d = 1030 \pm 130\pc, A_V = 1.1 \pm 0.46, R_V = 3.25\pm 0.65$ (see Fig \ref{distfig}).
These formal errors underestimate the true uncertainty, given the 
various assumptions that went into the model. Changes in the assumed
flux ratio between PDS 144S and 144N can shift the resulting best-fit distance
by an additional $\sim 100$ pc. Future work is certainly needed to pin down
the distance to PDS 144 more precisely. 
Nonetheless, for the moment we will
adopt $1000 \pm 200$ pc as a working estimate.

A major assumption in this distance fit is that PDS 144S has 
luminosity for an A5 star on the ZAMS. Many Herbig Ae/Be stars have
luminosities a few times greater than ZAMS
values. If PDS 144 is significantly pre-main-sequence and thus more
luminous than we assume, it will be correspondingly more
distant, which will only exacerbate the problem of its height above
the galactic plane (see below).

Another way to estimate the distance to PDS 144 is by comparison with
other Herbig Ae stars of known distance and similar spectral type. This has the advantage of
not requiring us to assume a particular luminosity for PDS 144S.
For instance, IP Per is a A6
star with V=10.5 at a distance of 350 pc \citep{2004AJ....127.1682H}. 
For it to have the same apparent brightness as PDS 144S, IP Per would
need to be located at 1200 pc, assuming the extinction is unchanged. 
Different results are possible depending on what star is picked as the 
analog of PDS 144S. Using a sample of ten A4-A6 HAes taken from the literature,
this procedure results in distances of 550-1900 pc. It is encouraging
that our 1000 pc estimate falls in the middle of this range, but these
comparisons do not reduce our uncertainty in the
distance to PDS 144.

These data seem to rule out any association of PDS 144 with Upper Sco at
140 pc.
Even with extraordinarily grey dust ($R_V > 20$),   
models of PDS 144S with a distance of 140 pc and sufficient extinction to produce
$V=13.1$ predict $V-R \approx 1$, inconsistent with the observed
$V-R=0.30$. Known Herbig Ae stars at the distance of Upper Sco are
either some 3-5 mag. brighter at $V$ than PDS 144 (e.g. KK Oph, HD
163296, or TY CrA) or are embedded objects much redder than PDS 144
(e.g. Elias 1, which has $V-R = 2.1$; \citet{1999AJ....118.1043H})
A larger distance also provides a natural explanation for the low
proper motion of PDS 144, which at $2\pm4$ mas/year is an order of magnitude slower than the
average for Upper Sco/Oph \citep{teixeira2000,2005A&A...438..769D}.

But a 1 kpc distance is not without problems of its own. Chief among
these is how did such a young system come to be so far above the
Galactic plane? At a Galactic latitude of +21\degr and a distance of
$1000\pm200$ pc, PDS 144 is $360\pm70$ pc above the Galactic plane.
\citet{kroupa1998} claims that stars of all masses can be ejected at
high velocities from moderate to dense clusters, so perhaps PDS 144 is
a runaway from a star forming region closer to the plane. But if that
is the case, it's difficult to understand how the binary would have remained
bound, since the necessary velocity to reach its current position in
less than a few million years (50-100 km/s) is greatly in excess of
the binary orbital velocity ($\approx 1$ km/s). Further contradicting
this scenario is that its proper motion ($2\pm4$ mas/year \citep{2005A&A...438..769D},
corresponding to a transverse velocity of 9 km/s at 1000 pc) is both slow and directed mostly \textit{towards}
the Galactic plane.  We are left with the conclusion that this system
must have formed far above the Galactic plane.  
It is unknown what star forming region may have given birth to this system;  
no dark clouds are found within 2 degrees of PDS 144 according to the catalog of \citet{2005PASJ...57S...1D}.

One possible resolution to this dilemma would be if we have entirely
misinterpreted the nature of these stars, and that they are in fact
evolved post-main-sequence objects rather than young stars.
Proto-planetary nebulae appear similar to pre-main-sequence A stars in
several aspects such as IR excess from dust, the presence of PAH
emission, and optical forbidden lines such as [SII].
The PDS
catalog is known to be contaminated by such interlopers
\citep[for instance, PDS 461/Hen 3-1475;][]{2002A&A...387..955G}.
There are two reasons we reject this interpretation: First, 
the morphology of PDS 144N more closely resembles edge-on disk systems
than it does proto-planetary nebulae such as PDS 461. Secondly,
\textit{both} stars in PDS 144 display IR excess and optical forbidden
lines. Since their spectral types and presumably masses differ, we
would not expect them to both evolve off the main sequence
simultaneously.  It is much more plausible for the pair to be young stars
approaching the main sequence together rather than departing it.

PDS 144 is by no means the only Herbig Ae/Be star whose distance has
proved problematic. \LkHa 101, long believed to be located at 800 pc,
may instead lie only 140 pc distant
\citep{1998AJ....116..890S}.  Conversely, the distance to HD 34282
has recently been revised upwards from 
160 to 350 pc \citep{2003A&A...398..565P,2004A&A...419..301M}. Similar
methods to those used in the just-cited papers may help clarify the
distance to PDS 144. In particular, high-resolution optical
spectroscopy could constrain stellar properties directly, while
millimeter observations of Keplerian rotation in the disk of PDS 144N
could provide a dynamical distance.

\section{Disk morphology }

\subsection{Basic Appearance}

At all wavelengths up to 11.7 $\micron$, the exciting star within PDS
144N is hidden from direct view by the dark lane of an edge-on
circumstellar disk.   The dark lane is $0\farcs15$ in height  and
extends $\sim 0\farcs8$ in the orthogonal direction (Fig \ref{keckfig}). Bright
scattered-light nebulosity is visible on both sides of the disk plane,
with the southeast side being brighter and therefore nearer, assuming
the grains are forward scattering.

At our assumed distance, the $0\farcs8$ apparent disk diameter
corresponds to a physical diameter of $800\pm180$ AU .  This 
is comparable to disks around other Herbig Ae stars
such as HD 100546 (500 AU) or HD 163296 (900 AU)
\citep{2000ApJ...544..895G,2001AJ....122.3396G, grady2005}.  However,
it is possible that the observed dark lane is actually a shadow cast
by a smaller disk into a surrounding envelope
\citep{2005A&A...435..595P}.  In this case the actual disk size could
be smaller than the 800 AU apparent size. Pontoppidan et al. list a
number of observational criteria which can be used to discriminate
disk shadows from large disks, such as the presence of a bright point
source in the middle of the dark lane or a ``wedge-like'' shape, but
none of these criteria fit PDS 144N.

PDS 144N is laterally asymmetric.  In the inner region of the nebula,
the peak intensity is displaced $0\farcs05$ southwest from the
vertical axis.  In the outer nebula, the northeast side
is more extended than the southwest.   Similar asymmetries
in the disk of HH 30 proved to be variable over time scales of a few
years, and have been
interpreted as changing illumination caused by the orbital
motion of clumps at a few AU \citep{1999ApJ...516L..95S}.
Alternatively, those asymmetries may be evidence for a hot spot on the
rotating stellar surface \citep{cotera2001}. 
Long term monitoring of
PDS 144N is needed to see whether its lateral asymmetries also
vary with time. If so, then the timescales and amplitude of the
variability may clarify whether it is driven by an outflow (such as by
a precessing jet), a hot spot, shadowing by orbiting dust clumps, or
some other mechanism. Alternatively, the asymmetry in PDS 144N could be due to
disk inhomogeneities at large radii where orbital periods are
many decades, and so we may never see any changes.

\subsection{The ``wings''}

One of the most striking features of the nebulosity around PDS 144N is
the pronounced flaring at its outside edges: it resembles  an
opposing pair of wingnuts, with the outer vertical extensions being the
``wings''. The wings begin approximately 0\farcs15 on either side of the
disk's vertical axis, and extend 0\farcs4 above and below the disk
midplane, twice the extent of the scattered light near the axis. The
wings are comparatively redder than the inner portions of the nebula;
they  are most pronounced in the PAH and $L'$ filters, but are still
visible at a lower level in $H$ and $K'$ (Fig. \ref{diskfig}). 

\citet{1994ApJ...428..654H} proposed that massive stars can launch
disk winds by photoevaporating the outer portions of circumstellar disks. Could such winds
boiling off the disk be responsible for creating the wingnut
morphology?  The predicted wind emerges from the outer disk along
outward-curving streamlines, which resemble the wings around PDS 144.
The characteristic radius at which such evaporation is expected to
take place is $    r_{evap} = GM_*/a_s^2 \simeq  7~\mathrm{AU}
(M_*/M_\odot) $, where $a_s$ is the sound speed in the disk.  But at a 1 kpc distance,  $r_{evap} \approx 0\farcs02$, 
an order of magnitude smaller than the observed size, and we must
find some other explanation for the wings.  

Perhaps the YSO that most resembles PDS 144N is \object{IRAS
04302+2247} \citep{padgett1999}, sometimes called the Butterfly Star
in Taurus. Like PDS 144N, it has a dark lane bisecting two bright
regions of nebulosity, each of which has its center fainter than its
outer edges giving an overall quadripolar morphology. Foreground
extinction has been suggested as the cause of IRAS 04302+2247's
quadripolar appearance, but we can rule this explanation out for PDS
144: the quadripolar appearance is still prominent at 11.7 microns
despite the much greater dust-penetrating ability of photons at that
wavelength.  \citet{Wolf2003} modeled IRAS 04302+2247 as a disk plus
an infalling circumstellar envelope with bipolar outflow cavities.
The cavity walls scatter more light than the evaculated cavities
themselves, providing a natural explanation for the ``wings'' of the
``butterfly'' - and perhaps also the ``wings'' of PDS 144N. We present
disk plus envelope models of PDS 144N in section 5. 

The wings become redder as one moves further from the disk midplane.
Intriguingly, this is opposite the blue color seen for the
envelope around HD 100546 by \citet{2005prpl.conf.8511A}.
Possible explanations for the red color include, in decreasing order
of likeliness, (1) The illuminating light scattered there is redder,
due to greater contribution from the hot inner disk: Grains in the
envelope see a larger solid angle subtended by the inner disk than do
grains nearer the disk midplane.  (2) Small grains in the envelope
emit thermal radiation after being transiently heated to $\sim600$ K.
(3) Dust grains in the envelope might instead be larger than disk grains
and thus scatter red light more efficiently; this seems unlikely as it is
inconsistent with our understanding of grain growth. (4) Scattered light from
the envelope may experience more extinction than that from the disk.
Some combination of factors 1 and 2 is most likely; these could be
distinguished by future $L'$ polarimetry observations.

\subsection{Disk Truncation?}

Like HV Tau C, PDS 144N's apparent diameter does not change significantly
with wavelength.  Since optical depth is a strong function of
wavelength, this suggests that the circumstellar material around
PDS 144N has a sharp outer edge, possibly because of tidal
truncation by PDS 144S.  The expected limit on disk size from tidal
truncation is roughly 1/3 of the binary's separation
\citep{1994ApJ...421..651A}. If we assume that PDS 144's orbit is
circular with the observed separation, 5400 AU, corresponding to the
major axis, then the expected tidal truncation radius is $\approx
1800$ AU, several times the observed radius. This is a lower limit on
the truncation radius, since the true separation must be greater than
or equal to the observed separation in the plane of the sky. Thus
while PDS
144 is not inconsistent with tidal interaction models, tides do
not seem to be the limiting factor on disk size here. 

However, a third source, such as a brown dwarf companion, might provide an 
alternative source of tides. We reiterate that there is a third, much fainter
source located almost directly between the two PDS 144 stars (Table 3). Future
observations are required to tell if this is a comoving member of the system
or not. But given the hierarchical nature of most
binaries and its position equidistant from both A stars, it is most
likely merely a background star.

There is another possibility: the radiative effects of PDS 144S may be
more important in truncating the disk than its gravity. Far UV
photons from PDS 144S could heat and evaporate the outer disk, as has
been suggested to occur for disks around low-mass stars externally
illuminated by nearby massive stars, \citep{2004ApJ...611..360A}.
However, unlike typical proplyds in Orion where the external
illuminating source is much brighter than the stars within the disks,
in PDS 144 both sources are of comparable intrinsic luminosity.  More
detailed modeling is needed to disintangle the photoevaporative
effects of the two stars. 

We caution that the truncation of PDS 144N's disk and envelope by either tidal or
photoevaporative forces is speculative. The presence of substantial IR excesses
around both stars and the lack of any visible tidal streamers or
other debris from a disrupted disk suggest there is an alternative argument to be
made for little interaction between the two stars.

\subsection{PAH emission}

Polycyclic aromatic hydrocarbons (PAHs) produce several IR emission features frequently seen
around Herbig Ae stars \citep{1989A&A...216..148L,1995A&A...302..849N}. 
These emission bands arise due to fluorescent excitation of PAHs by far-UV
photons, which are produced in great quantity by hot A stars.

PDS 144N has emission several times brighter than continuum levels in the
3.3 ${\mu}m$ PAH band (Fig. \ref{diskfig}).
We used for the 3.3 $\micron$ continuum a linear interpolation between the 
2.1 $\micron~K'$ and $3.8~\micron~L'$ filters, neither of which
contain the PAH feature within their bandpasses.  
The PAH emission is detected over
roughly the same region as is the continuum, but their relative
intensities vary. As measured in NIRC2's 0.05-$\micron$-wide PAH
filter, the PAH emission is $\sim 6$ times brighter than continuum in
the outer nebula, but in the bright core nearest the central star, the
PAHs are only $2$ times the continuum. In other words, the PAH
emission is less centrally peaked than is the continuum.

In order for PAH emission to occur, the PAHs must be illuminated by
far-UV photons. Flared disks allow starlight to directly fall on PAHs
in the disk surface, favoring PAH emission, while flat disks tend to
lack PAH emission because PAHs are in the shadow of the inner
disk \citep{Acke2004,Habart2005}. PDS 144N is consistent with this scenario: 
the PAH emission we see comes primarily from particles in the envelope, 
high above the disk's midplane. The models of \citet{Habart2004} predict that the 3.3 $\micron$ PAH
emission will be strongest within 100 AU of the star, a conclusion
supported by recent observations. Here we detect
that feature at several times greater distance.  We find that the
PAH/continuum ratio increases with distance
from the star, as predicted by Habart et al. However, the models of
Habart et al. were computed for face-on inclinations, so we cannot
directly compare them with PDS 144N's edge-on geometry.

The Spitzer IRAC 8 $\micron$ flux for PDS 144N
is significantly greater than the $11.7 \micron$ LWS flux. This
suggests that PDS 144N has strong PAH emission in the 7.7 or 8.6
$\micron$ PAH bands as well, but higher spectral resolution
observations are needed to confirm this.

\subsection{The Mid Infrared}

What is the origin of the photons we detect from PDS 144N at 11.7
$\micron$?  The outer disk is too cool to emit thermal photons
over the observed spatial extent.  Nor can the light be scattered
stellar photons, since the observed flux is far in excess of the
photospheric value at this wavelength.  However, it may be scattered
photons from the hot inner disk, as is seen around HK Tau B at this
wavelength \citep{2003ApJ...588L.113M}. PDS 144N must have substantial
thermal emission from its inner disk (unless it has a large inner
radius, which we cannot exclude based on our data) providing
sufficient photons to scatter and produce the observed mid-IR
nebulosity.  Alternatively, it could be emission from the 11.3
$\micron$ PAH band, which falls within the bandpass of LWS's 11.7
$\micron$ filter.  The strength of the 11.3 $\micron$ PAH feature
should correlate with the 3.3 $\micron$ feature, since both bands
arise from the same aromatic C-H bonds \citep{1994ApJ...423..326B},
and we know there is copious 3.3 $\micron$ PAH emission present.  The
energetics are favorable for PAH emission, too: A main-sequence A2V
star supplies sufficient UV flux to energize PAHs enough to produce
the 11.7 $\micron$ light, even before adding any UV excess from active
accretion.  With these data we cannot conclusively resolve this issue.
In order to further constrain the emission mechanism, we have obtained
additional observations in a series of narrowband mid-IR filters with
Michelle at Gemini, which will be presented in an upcoming paper
(Perrin et al., in preparation).

\subsection{PDS 144 South}

\citet{vieira2003}'s optical spectroscopy confirmed the presence of
emission lines in the spectrum of PDS 144S, including [SII] $\lambda
\lambda$ 6717, 6730 and [OII] $\lambda\lambda$ 3726, 3728 in addition
to H$\alpha$. Therefore PDS 144S is actively accreting and must
necessarily possess circumstellar material of its own. Furthermore, we
find PDS 144S has almost 5 magnitudes of IR excess at $K_s$ band,
unambiguously indicating the presence of warm circumstellar dust. However,
the SED alone cannot tell us the distribution of that dust; models
using either disks or spherical envelopes can both adequately reproduce dusty
SEDs. Resolved imaging is necessary to break the degeneracies between
model parameters \citep{chiang2001}.

We do not resolve a disk around PDS 144S in any of these observations
(though we remind the reader that our Keck data is saturated on PDS
144S in all but one waveband, preventing PSF subtraction and
greatly degrading disk detection limits).  
Given PDS 144S's unresolved appearance, any disk around PDS 144S must
be either noncoplanar with PDS 144N's disk or sufficiently less
massive that it is optically thin at these wavelengths, so that the
system's appearance is dominated by the direct light from the star.
\citet{2004ApJ...600..789J} observed 19 multiple T Tauri systems and
found that 8 out of 9 binaries had polarization vectors aligned within
30\degr, consistent with disk alignment. However, these binaries have
apparent separations 200-1000 AU, much less than PDS 144's 5400 AU. It is
plausible that disk alignment mechanisms may be less effective in
systems with greater separations. HK Tau provides a
well-known example of non-coplanar disks in a system apparently quite
similar to PDS 144 \citep{1998ApJ...502L..65S,Duchene2003HKTau}. 

Unlike PDS 144N, PDS 144S displays no evidence for PAH emission above continuum levels
in the present observations;  the difference in the shapes of the two
stars' SEDs from 3.6-8 $\micron$ is quite striking. PDS 144S has
sufficient far-UV flux to energize PAHs if present. Possibly the two
stars' disks differ in composition, with PDS 144S lacking PAHs. Alternatively, \citep{Acke2004}
suggests that Herbig Ae stars with flat disks have much
weaker PAH emission than stars with flared disks because
self-shadowing prevents UV light from reaching the PAHs. Thus the two
PDS 144 stars could have similar abundances of PAHs yet very different
spectra if their circumstellar dust geometries differ.

\section{Preliminary Disk Models}

We have modeled the disk around PDS 144N using the Monte Carlo
radiative transfer code made available by B. Whitney and collaborators
\citep{2003ApJ...591.1049W,2003ApJ...598.1079W}.  This was a
first-draft modeling effort only, with model parameters tuned by eye
to approximate the observed properties of PDS 144 at $H$ band.  No
attempt has been made here to minimize errors in any formal
statistical sense.  The purpose of this exercise was to 
identify a reasonable starting point for future more detailed
modeling, and to determine which if any features of PDS 144N are most
discrepant from simple disk models.  A more sophisticated fit to the
complete multiwavelength data set via a multidimensional optimization
code will be described in a future work. Here we shall limit
ourselves to discussing the overall properties of our model, and the
most notable discrepancies between model and observation.

The Whitney et al. radiative transfer code allows grain properties to vary between
different regions. Briefly, grains are treated as spheres with scattering computed using
the Mie theory; grains follow power-law size distributions with
parameters varying from ISM-like in the envelope and outflow
to larger grains in the disk and midplane.  
We refer readers to  
\citet{2003ApJ...598.1079W} for a detailed discussion of assumed grain
parameters; we used an identical prescription as those authors.
We assumed a disk mass of 0.01 $M_\odot$, outer radius $r=350$ AU, density exponent
$\alpha = 2.25$, scale height $h=0.4$ AU at $r=50$ AU, and scale height exponent $\beta = 1.25$.

We first attempted to model PDS 144N using the simplest possible model
of only a passive circumstellar disk, without any surrounding
envelope. We did this computation for $H$ band,
the wavelength in which the ``wings'' are least pronounced.  The
overall shape of the disk, including the slight curvature of the dark
lane, is best reproduced by disk models with inclinations of $83 \pm 1
\degr$.  Despite an extensive search of parameter space, the simple
disk models were unable to reproduce the extended nebulosity around
the disk.  Furthermore, even though the wings are minimized at $H$
band, the northwestern lobe still has a local minimum on the vertical axis,
while the disk-only models predict a local maximum there.  Thus we
conclude that a circumstellar envelope is required in addition to a
disk to accurately model this system.

\begin{figure}[!ht]
\begin{center}
\plotone{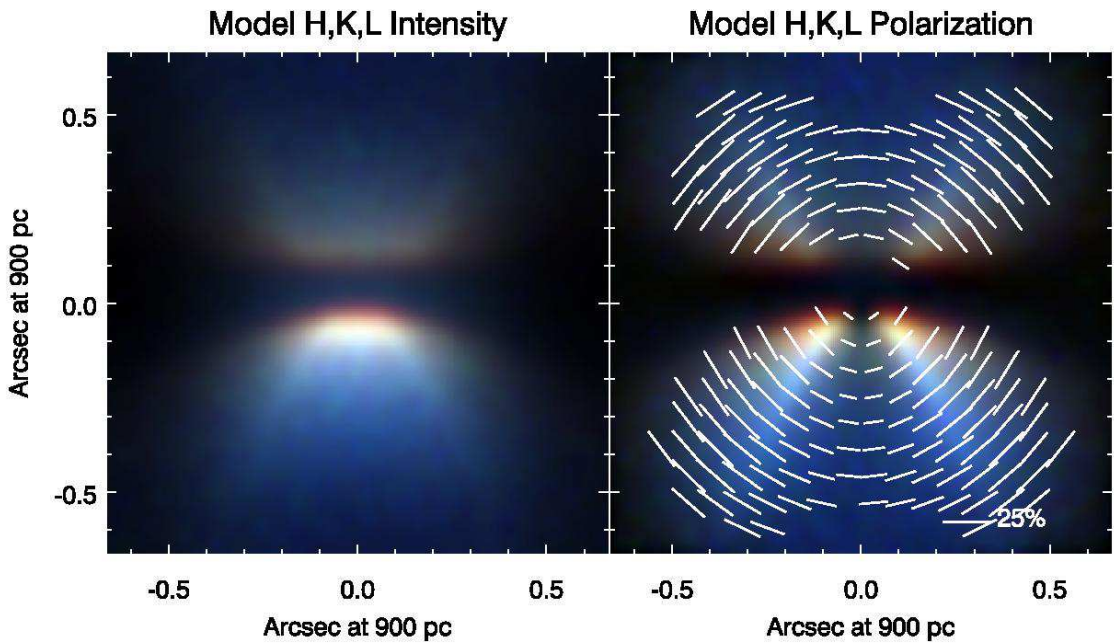}
\end{center}
\vskip -0.30in
\caption{
\footnotesize
\label{modelfig}
Monte Carlo model of a circumstellar disk plus envelope, at 1000 pc,
convolved with a wavelength-dependent model Keck PSF computed with the Keck AO model
of Sheehy et al. (in preparation). The left frame shows total
intensity $I$ while the right frame shows polarized intensity
$\sqrt{Q^2+U^2}$. $H, K'$, and $L'$ are blue, green, and red,
respectively, and the image scale is linear. Vectors in
the right hand frame show the degree and orientation of $K'$ band
polarization.
The disk model 
has mass 0.01 $M_\odot$ and radius 350 AU. Surrounding that is a
rotationally supported infalling envelope \citep{1984ApJ...286..529T}
with an infall rate of $2\times10^{-6} M_\odot \yr^{-1}$ and a cavity opening
half angle of 15\degr. 
}
\end{figure}

Models incorporating a circumstellar envelope and outflow cavities are
much more successful in reproducing the observed morphology.  We added
a rotationally supported infalling envelope
\citep{1984ApJ...286..529T} with an infall rate of $2\times10^{-6}
M_\odot \yr^{-1}$ and a cavity opening half angle of 15\degr.  We
display this model in Figure \ref{modelfig}. Limb-brightening from the
edges of the outflow cavity roughly reproduces the ``wings'' seen
around PDS 144N. 

However, there remain some discrepancies between this model and
the data. While this is only a preliminary model, we point out some 
differences so that they may be addressed in future more comprehensive
modeling.
The observed nebula has a sharper outer edge than the model does. This
is not a signal-to-noise ratio effect: the observed nebula outer radius is
roughly the same at all wavelengths, even though the SNR varies
substantially. Furthermore, introducing read- and sky-subtraction
noise comparable to the observed levels into our model images does not
truncate the apparent size of the disk.  We have discussed above possible
scenarios for tidal or photoevaporative truncation of the material
around PDS 144N. 
Also, the model predicts that for an inclination of 83\degr, the nearer side
of the disk (southwest) will have a peak surface brightness $\sim
3$ times the fainter side.  The observed southwest side is
actually 10 times brighter than the fainter side.  This difference may
reflect the model's use of Mie scattering by spherical grains
rather than more realistic grain scattering.

\section{Conclusions}

We have presented high resolution observations of PDS 144 from 1.2-11.7 $\micron$
obtained at Lick and Keck Observatories, plus archival Spitzer
photometry from 3.6-70 $\micron$. 
PDS 144N is hidden from direct view at all wavelengths from 1.2-11.7
$\micron$ by an edge-on optically thick circumstellar disk $0\farcs8$
across. Red ``wings'' extend vertically from the disk, most likely
due to scattering from the walls of an outflow cavity in an infalling
envelope of dust and gas. The wings are very bright in the 3.3
$\micron$ PAH feature, indicating the presence of small aromatic
grains fluorescing due to far-UV illumination. Its nearby companion, PDS 144S, is unresolved
in our data but has very strong IR excesses indicating that it, too,
is surrounded by copius circumstellar dust. But a number of basic questions remain open about this fascinating
system:

How far away is it? Our distance estimate of $1000\pm200$ pc is
based on fitting the $UBVRI$ measurements of \citet{vieira2003} with
an extincted A5 stellar spectrum for PDS 144S. Uncertainties in the 
spectral type and evolutionary state of the stars, plus confusion of the two sources in the unresolved
optical photometry, limit the precision possible with this method. The
1000 pc distance also results in a surprising height above the
Galactic plane. We caution that this distance estimate relies on the assumption that the
two stars of PDS 144 are physically associated, which is probable
but unproven.

How old is it? The uncertain distance results in similarly uncertain total luminosities
for the two stars, making it difficult to place them on
pre-main-sequence evolutionary tracks \citep[e.g.][]{1993ApJ...418..414P}. 
The large IR excesses of both stars suggest there may be significant accretion luminosity. This further
complicates placing the stars on pre-main-sequence tracks, but also implies the system
is quite young. 

How massive is it? The stellar masses are poorly constrained for the same reason
their ages are, because their positions in the H-R diagram are imprecise. The
dust mass around PDS 144N cannot be constrained with the data presented here
since the disk is optically thick at these wavelengths.  Resolved millimeter
observations of this system would be very useful to constrain the dust masses
around each star. Given the southern declination, the SMA may be the array best
suited to obtaining these observations. 

How have the two stars interacted? PDS 144N's envelope appears to have
been truncated, perhaps due to the influence of PDS 144S, although 
the mechanism for this truncation remains unclear. The two
stars both have circumstellar material, but have very different
appearances at these wavelengths. Can this be explained entirely by geometric
effects due to different viewing angles to physically similar
young stars, or are the stars and their circumstellar material
intrinsically different? If so, why, given that they are presumably
coeval, of fairly similar mass, and formed from the same parent cloud?

Are there outflows? The presence of a bipolar cavity around PDS 144N
suggests that there are, or at least once were.
Polar outflows from disks around T Tauri stars may be traced by a number of
low-excitation emission lines, such as in the exquisite HST images
of the HH 30 disk and jet \citep{1996ApJ...473..437B}. 
Similar outflows have been seen around a number of Herbig Ae stars 
\citep{corc98,2000ApJ...544..895G}.
Does PDS 144N also harbor such jets?
The presence of [SII] emission in the optical suggests that it does. At
near IR wavelengths where AO imaging is possible, the [FeII] 1.257,1.644 $\mu$m lines are ideal
outflow tracers on account of their low ionization potential and
excitations  \citep{2003A&A...410..155P,2003ApJ...590..340P}.

As the first Herbig Ae star with a clearly resolved edge-on disk, PDS
144N provides an exciting opportunity for comparison of disk
properties between intermediate-mass and lower-mass stars, such as the
well-studied HH 30.  The geometry of the PDS 144 system makes it a
particularly well-suited laboratory for detailed studies of
circumstellar material.  Such studies need both high sensitivity and
high angular resolution.  Because the disk is edge-on, it blocks the
direct starlight, resulting in a greatly favorable contrast ratio and
thus increased sensitivity to faint emission.  Furthermore, since PDS
144 is a binary, we have a bright unocculted star conveniently nearby
to serve as an wavefront reference source. Just as in HK Tau, the
binarity of this system enables studies which would otherwise not be
possible.

\acknowledgments

We greatly appreciate the excellent work of the Lick and Keck 
observatory staffs, in particular Elinor Gates,
Mark Kassis, and Randy Campbell.
We also thank Bob Becker for his gracious flexibility in
trading LWS observing time, and a two dollar bill. Sergio Vieira and
Carlos Torres provided helpful discussions regarding their PDS survey
observations.
We recognize that
Mauna Kea has always been a site of great significance to the Hawaiian
people, and we are most fortunate to have the opportunity to observe from this
mountain.

    This work has been supported in part by the National Science
      Foundation Science and Technology Center for Adaptive Optics,
      managed by the University of California at Santa Cruz under
      cooperative agreement No. AST-9876783.
     PK received additional support from the NASA
      Origins Program under grant NAG5-11769.  MDP is supported by a
      NASA Michelson Graduate Fellowship, under contract to the Jet
      Propulsion Laboratory (JPL). JPL is managed for NASA by the
      California Institute of Technology.

{\it Facilities:} \facility{Shane} \facility{Keck:I}
\facility{Keck:II} \facility{Spitzer}

\clearpage

\bibliographystyle{apj}
\bibliography{haebes}

\end{document}